\documentclass[12pt,twoside]{article}
\usepackage{fleqn,espcrc1}

\usepackage{graphicx}
\usepackage[figuresright]{rotating}

\newcommand{\AmS}{{\protect\the\textfont2
  A\kern-.1667em\lower.5ex\hbox{M}\kern-.125emS}}
\newcommand{\be}{\begin{equation}}
\newcommand{\ee}{\end{equation}}
\newcommand{\bea}{\begin{eqnarray}}
\newcommand{\eea}{\end{eqnarray}}
\newcommand{\psat}{p_{\rm sat}}
\renewcommand{\d}{{{\rm d}}}

\newcommand{\lton}{\mathrel{\lower.9ex
                  \hbox{$\stackrel{\displaystyle <}{\sim}$}}}

\title{The Transverse Energy as a Barometer of a Saturated Plasma}

\author{A.\ Dumitru$^{\rm a}$ and M.\ Gyulassy\address{Physics Department, Columbia
Univ., 538 W. 120th Street, New York, NY 10027, USA}}
       
\begin{document}

\maketitle

\begin{abstract}
The evolution of the gluon plasma produced with saturation initial conditions
is calculated via Boltzmann transport theory for nuclear collisions at high
energy. The saturation scale increases with $A$ and $\surd{s}$, and thus
we find that the perturbative rescattering rate decreases relative to the
initial longitudinal expansion rate of the plasma.
The effective longitudinal pressure remains
significantly below the lattice QCD pressure
until the plasma cools to near the confinement scale.
Therefore, the transverse energy per unit of rapidity and its dependence on
beam energy provides a sensitive test of gluon saturation models:
the fractional transverse energy loss due to final state interactions
is smaller and exhibits a weaker energy dependence
than if ideal (nondissipative) hydrodynamics applied throughout the evolution.
\end{abstract}

In collisions of heavy ions at high energy a large number of gluons is
liberated from the nuclear wave functions.
The ``plasma'' is produced from copious minijet gluons at central rapidity,
$y\simeq0$, with transverse momentum $p_T>p_0$.
For large $p_0$, the produced gluon plasma is dilute. As
$p_0$ decreases, however, the density of gluons
increases rapidly due to the increase
of $G(x,p_T^2)$ as $x\approx 2p_T/\surd s$ decreases. It has been
conjectured~\cite{glr,McLerran:1994ka} 
 that below  some transverse
momentum scale $p_0\le \psat$ the phase-space density of produced gluons
may saturate since $gg\rightarrow g$
recombination could limit further growth of the structure functions.
Phenomenologically, this condition may arise when
gluons (per unit rapidity
and  transverse area) become closely packed and
 fill the available nuclear interaction transverse area.
The saturation scale $\psat$ can  thus be estimated from
${\d N}(\psat)/{\d y}=\psat^2 R_A^2/\beta$,
where $\beta\sim 1$.
For $\beta=1$ the solution reported  in EKRT~\cite{kimmo}
was $\psat \approx 0.208A^{\; 0.128}\,\sqrt{s}^{\; 0.191}$,
where $\psat$ and $\sqrt{s}$ are in units of GeV, and
$C_1 \equiv \d E_T/\d y/\psat\d N/\d y
 \approx 1.34A^{-0.007}\,\sqrt{s}^{0.021}$. Our focus
is to investigate  whether the final observed $\d E_T^f/\d y$ can be used
to test the predicted $A$ and $\surd s$ dependence of
the  initial  $\d E_T^i/\d y$.

Different gluon saturation models based on classical Yang-Mills
equations~\cite{McLerran:1994ka,Mueller:2000fp} suggest that
the factor $\beta$ may vary parametrically as
$\beta(\psat) = {4\pi\alpha(\psat)N_c}/{c (N_c^2-1)}$,
where $c\sim 1$ is a nonperturbative factor proportional to the
fraction of the initial gluons in the nucleus which are liberated.
This factor was recently estimated~\cite{KrVe} to be $c\approx 1.2 - 1.5$.
On the other hand, PHOBOS and PHENIX data~\cite{phobos} require
$c\approx 1.9$ if one assumes entropy conserving hadronization,
and for $\alpha$ running with $\psat$; that is,
$\sim50\%$ additional
entropy must be produced during nonideal expansion of the saturated gluon
plasma. In the approach of~\cite{son}, the ratio of final to initial
multiplicity (or entropy) also grows with energy as $\sim\alpha^{-2/5}$.
This, in turn, suggests that less mechanical work is being performed
by the longitudinal expansion than predicted by ideal fluid
dynamics~\cite{admik}, and
so the $E_t$ in the final state should be closer to it's initial value.

If the expansion proceeds in approximate local equilibrium with
pressure $p= c^2 \epsilon$ and speed of sound $c$, then 
the energy density, $\epsilon(\tau)$, must decrease faster than the expansion
rate $\Gamma_{exp}=1/\tau$ and
leads to a  bulk  transverse energy loss
\be \label{edens_ansatz}
\frac{E_T(\tau)}{E_T(\tau_0)} =\frac{\tau\epsilon}{\tau_0\epsilon_0}= 
\left(\frac{\tau_0}{\tau}
\right)^\delta\quad.
\ee
{\em If} local equilibrium
is maintained during the evolution $\delta=c^2$. In contrast, if the system
expands too rapidly to maintain local equilibrium, then
the effective pressure is reduced due to dissipation.
The extreme asymptotically free plasma case corresponds
to free streaming with
$\delta=0$. $E_T$ thus provides an important
 barometric observable
that  probes the (longitudinal) pressure in the  plasma.

The relevant relaxation rate is given by the fractional 
energy loss per unit length, $\Gamma_{rel} = {\d \log E}/{\d z}$,
which receives a contribution both from elastic and inelastic scattering.
The relaxation rate is approximately given by~\cite{admik}
\be \label{relax_rate}
\Gamma_{rel} \approx 9\pi\alpha^2 \frac{\rho^3}{\epsilon^2}
\left(\log\frac{1}{\alpha} + \frac{27}{2\pi^2}\right)
\equiv K_{in} 9\pi\alpha^2 \frac{\rho^3}{\epsilon^2}
\log\frac{1}{\alpha}~,
\ee
where we lumped energy loss from radiation into $K_{in}$, the inelasticity
$K$-factor. $K_{in}=1$ corresponds to purely elastic scattering,
while $K_{in}=2$, for example, corresponds to twice the
lowest-order elastic scattering rate.

At the initial time $\tau_0=1/\psat$, $\overline{s}/2=\epsilon_0^2/
\rho_0^2=C_1^2\psat^2$.
Therefore, noting that the comoving gluon density at
time $\tau_0$ is $\rho_0=\psat^3/\pi \beta$, the ratio of
the relaxation  rate to the expansion rate is given by
\be \label{scatt_to_exp}
\frac{\Gamma_{rel}}{\Gamma_{exp}}=
 K_{in} \frac{9\alpha^2}{\beta C_1^2}
 \log\frac{1}{\alpha}\quad.
\ee
While $\Gamma_{rel}\propto \psat$ increases  as a power of the energy,
the Bjorken boundary conditions
force the system to expand londitudinally initially also at 
an increasing rate $\Gamma_{exp}(\tau_0)=\psat$. 
The essential quantity that fixes the magnitude of the effective
pressure relative to that predicted by lattice QCD (LQCD)
is the ratio of rates in Eq.~(\ref{scatt_to_exp}), which dimensionally
is simply a function of $\alpha(\psat)$.
The asymptotic freedom property of QCD therefore
requires that this ratio vanishes as $\surd{s}\rightarrow \infty$.
In~(\ref{scatt_to_exp}) the rate of how fast it vanishes is controlled by 
$K_{in}\alpha^2 (\psat)/\beta(\psat)$. Therefore, with
saturation initial conditions, asymptotic freedom reduces the {\em effective}
pressure acting at early times
$\tau\sim \tau_0$ and causes the initial evolution to deviate
from ideal hydrodynamics for a time  interval  that
increases with energy~\cite{admik}. 

For a more quantitative estimate of the
transverse energy loss due to longitudinal work, we  employ the
Boltzmann equation in relaxation time approximation~\cite{Baym}.
As initial condition, we assume that the initially produced
partons have a vanishing longitudinal momentum spread in the comoving frame,
i.e.\ we assume a strong correlation between space-time rapidity and
momentum space rapidity; the accuracy of this approximation
at the finite RHIC energy remains to be checked within more elaborate
models. In any case, our gluon plasma thus starts at zero longitudinal
pressure, and we shall follow its evolution up to the point where
it reaches the $p/\epsilon$-ratio from LQCD~\cite{admik}.
\begin{figure}[htb]
\begin{minipage}[t]{75mm}
\includegraphics[width=65mm]{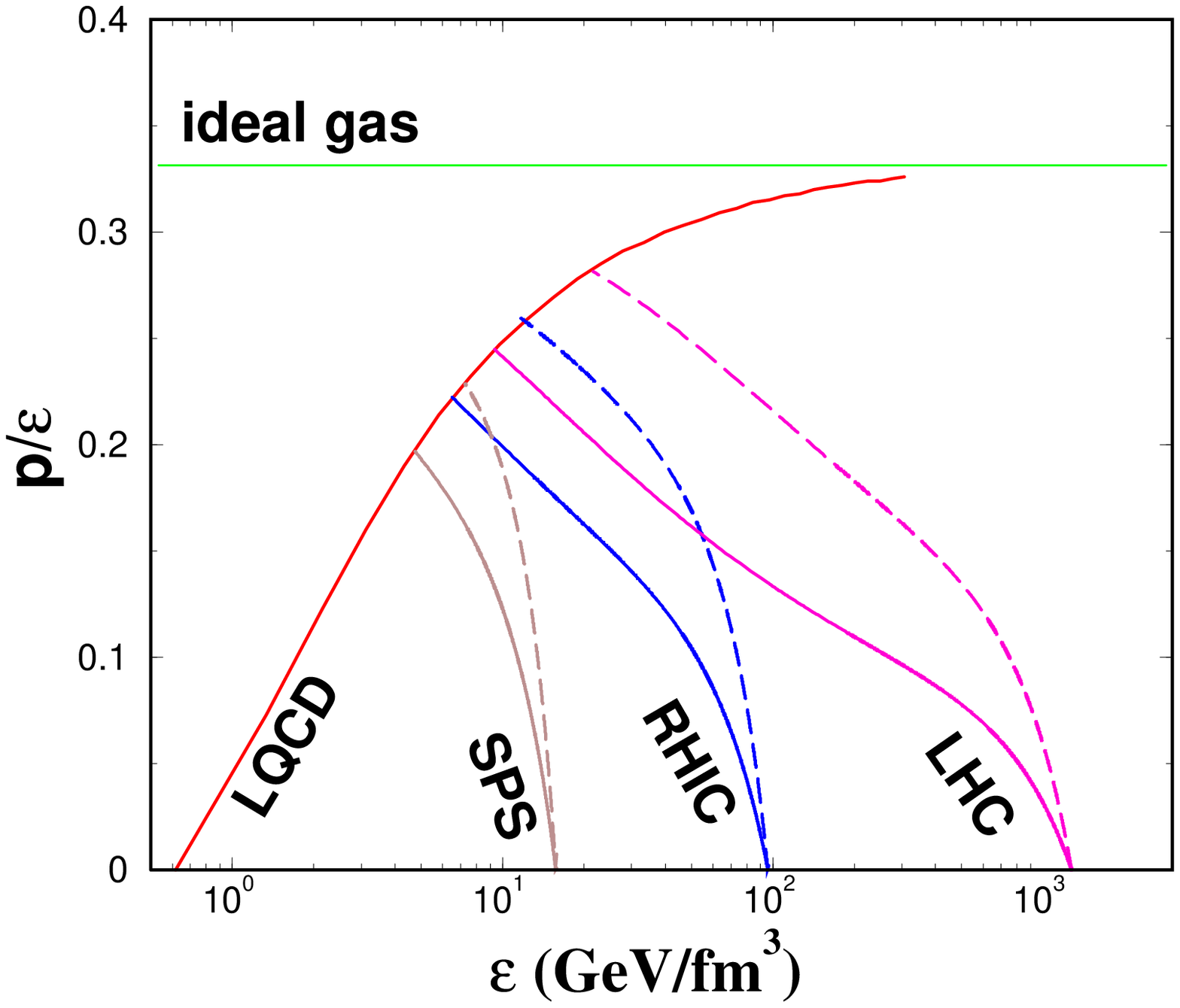}
\vspace{-10mm}
\caption{The ratio of the effective longitudinal pressure to the energy density
along the dynamical path is shown
for SPS, RHIC, and LHC saturation initial conditions\protect\cite{kimmo}.
Solid (dashed) curves are for $\beta=1$ and $K_{in}=1(2)$.
The LQCD equation of state~\protect\cite{lattice}
is also shown for comparison.}
\label{poe}
\end{minipage}
\hspace{\fill}
\begin{minipage}[t]{75mm}
\includegraphics[width=66mm]{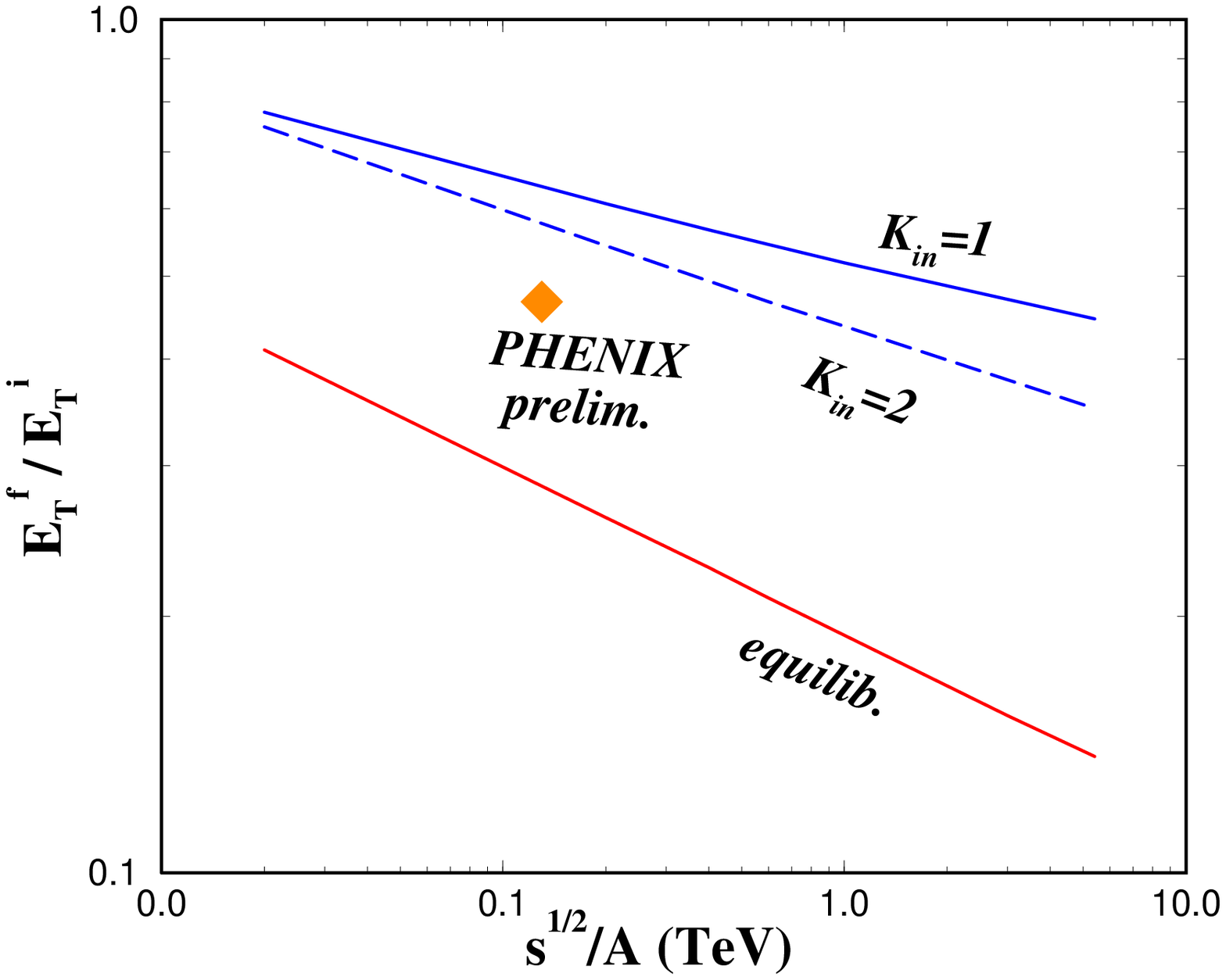}
\vspace{-10mm}
\caption{The ratios of the final to the initial
transverse energy per unit rapidity
are shown as a function of beam energy for central
Au+Au collisions. The initial value corresponds to the EKRT
parametrization ($\beta=1$) and the transport results are for $K_{in}=1,2$.
The final transverse energy for an ideal gas of gluons in
equilibrium (throughout the evolution) is also shown.}
\label{et}
\end{minipage}
\end{figure}
In Fig.~\ref{poe} we show the effective
longitudinal pressure as a function of the
energy density for $\surd s=20, 200, 5400A$~GeV saturation
initial conditions~\cite{admik}.
Initially $p/\epsilon$ starts at zero
and remains small for a large time relative to $\tau_0=1/\psat$ 
because the plasma is torn apart by the initial
rapid longitudinal expansion.
The effective pressure 
approaches the LQCD curve from below and reaches it at a time
$\tau_L\approx 1-2$~fm at RHIC, $\surd {s}=200A$~GeV, by which
time the energy density 
has dropped by an order of magnitude, $\epsilon_L\equiv\epsilon(\tau_L)
=6.5-12$~GeV/fm$^3$.
For LHC $\surd {s}=5400A$~GeV,  $\tau_L=3-7$~fm during which
the energy density falls by almost two orders of magnitude to $\epsilon_L=
9.5-21.5$~GeV/fm$^3$. The quoted intervals correspond to $K_{in}=1-2$
using the EKRT parametrization~\cite{kimmo}.
The contrast between the dynamical path followed by the saturated plasma
compared to the equilibrium equation of state is striking.
A qualitatively similar behavior of the early
logitudinal pressure has also been found from solutions of
diffusion equations~\cite{Jeff}.

The main experimentally
observable consequence of the reduced effective pressure
is  shown in the right panel.
The ratio ${E_T^f}/{E_T^i} = 
{\tau_f\epsilon_f}/{\tau_0
\epsilon_0}$ has been obtained from the solution of the transport equation
assuming free streaming at $\epsilon_f=2$~GeV/fm$^3$ which corresponds
roughly to $T\simeq T_c$. $\tau_f$ is determined
assuming hydrodynamic expansion from the point were the LQCD pressure
curve is reached. On the other hand, for an ideal gas of gluons produced at
time $\tau_0$, the final observed transverse energy for 1+1 dimensional
adiabatic expansion would be
\be \label{adiab_exp}
\frac{E_T^f}{E_T^i} = \frac{\tau_f\epsilon_f}{\tau_0\epsilon_0}
= \frac{\tau_f(T_f s_f - p_f)}{\tau_0(T_0 s_0 - p_0)} = \frac{T_f}{T_0}.
\ee
Clearly, for $T_f\sim T_c$ one would observe a much smaller
transverse energy in the final state than in the initial state.
Moreover, $E_T^f/E_T^i$ would also have
significantly stronger energy dependence such that $E_T^f$ deviates
more and more from $E_T^i$ with increasing $\surd s$.
In this sense isentropic hydrodynamics
erases information on the interesting initial conditions via  this observable.
The solutions of the transport equations clearly show a smaller decrease of
$E_T^f$ and of the logarithmic slope, $\kappa=\d \log E_T^f/\d\log \surd s$,
due to final state interactions. We find that  $\kappa=0.50$ for the
evolution with $K_{in}=1$, $\kappa=0.46$ with  $K_{in}=2$, while 
$\kappa=0.40$ with isentropic  expansion, eq.~(\ref{adiab_exp}).
For comparison, the initial EKRT saturated
$E_T^i=\pi R_A^2 \tau_0\epsilon_0$ scales with the higher power
$\kappa=0.59$.
The fractional transverse energy loss is thus  less
dependent on energy than
for entropy conserving expansion for which
$E_T^f/E_T^i\propto 1/T_0\propto
1/\sqrt {s}^{\; 0.2}$.
This is due to the increasingly long time spent far from equilibrium in 
Fig.~\ref{poe} as the beam energy increases.

The results in Fig.~\ref{et} are 
encouraging from the point of view of searching for evidence 
of gluon saturation in nuclei at high energies.
Experimental data on $\d E_T/\d y$ or $\d E_T/\d \eta$
for central Au+Au collisions at RHIC will soon provide a new  test of
saturation and non-saturation models at those energies. Since
we predict that dissipative effects reduce considerably
the effective longitudinal pressure in Fig.~\ref{poe}, 
the beam-energy dependence of the transverse energy
is expected to reflect much more accurately 
the predicted power law dependence of the initial
conditions as seen in Fig.~\ref{et}.
We have also shown the preliminary result from PHENIX~\cite{PHENIXET},
$\d E_t/\d\eta\approx570$~GeV for the $2\%$ most central
Au+Au events at $\surd{s}=130A$~GeV, scaled to $\d E_t/\d y$ and
divided by the initial $E_t$ from EKRT.
It would be most interesting to have data both at higher and lower energy
to determine the slope $\kappa$, and compare to the expectation for a
saturated plasma.
The energy and $A$ systematics of
the bulk calorimetric observable,
$\d E_T/\d y$, will be a sensitive test of saturation models
of gluon plasmas produced in the RHIC to LHC energy range.\\
{\bf Acknowledgement:} We acknowledge support from the DOE Research Grant, 
Contract No.DE-FG-02-93ER-40764.

\end{document}